\documentstyle[12pt,epsf]{article}
\newcommand{\no}{\nonumber}
\newcommand{\R}{{\bf R}}
\newcommand{\C}{{\bf C}}
\newcommand{\cH}{{\cal H}}
\newcommand{\del}{\partial}
\newcommand{\na}{\nabla}
\newcommand{\bra}{\langle}
\newcommand{\ket}{\rangle}

\newcommand{\Tr}{{\rm Tr}\,}
  
\newcommand{\rar}{\rightarrow}

\newcommand{\fr}{\frac}
\newcommand{\half}{\frac{1}{2}}
\newcommand{\qua}{\frac{1}{4}}
\newcommand{\scr}{\scriptsize}  
\newcommand{\dis}{\displaystyle}
\newcommand{\mn}{{\mu\nu}} 
\newcommand{\zb}{\bar{z}}
\newcommand{\beq}{\begin{equation}}
\newcommand{\eeq}{\end{equation}}
\newcommand{\beqa}{\begin{eqnarray}}
\newcommand{\eeqa}{\end{eqnarray}}

\newcommand{\pa}{\partial}
\newcommand{\mh}{{\hat{\mu}}}

\makeatletter
\@addtoreset{equation}{section}

\makeatother

\begin{document}

\begin{titlepage}
\null
\begin{flushright}
UT-913
\\
hep-th/0010221
\\
October, 2000
\end{flushright}

\vskip 1.5cm
\begin{center}

  {\LARGE On Exact Noncommutative BPS Solitons}

\lineskip .75em
\vskip 2cm
\normalsize

  {\large Masashi Hamanaka\footnote{e-mail:
 hamanaka@hep-th.phys.s.u-tokyo.ac.jp} and Seiji Terashima
\footnote{e-mail: seiji@hep-th.phys.s.u-tokyo.ac.jp}}

\vskip 2cm

  {\large \it Department of Physics, University of Tokyo,\\
               Tokyo 113-0033, Japan}

\vskip 2cm

{\bf Abstract}

\end{center}

We construct 
new exact BPS solitons in various noncommutative gauge theories
by the ``gauge'' transformation of known BPS solitons.
This ``gauge'' transformation introduced by Harvey, Kraus and Larsen
adds localized solitons to the known soliton.
These solitons include, for example, the bound state of 
a noncommutative Abelian monopole and $N$ fluxons at threshold. 
This corresponds, in superstring theories, to 
a D-string which attaches to a D3-brane 
and $N$ D-strings which pierce the D3-brane, where all D-strings are
parallel to each other.

\end{titlepage}

\clearpage

\baselineskip 7mm


\section{Introduction}

In the past few years, 
there has been much development in our understanding
of various properties of noncommutative field theories.
In particular, it has been shown that
some noncommutative gauge theories can be embedded 
in string theories \cite{CoDaSc}-\cite{SeWi}.
This fact means that there exist consistent noncommutative field theories 
even quantum mechanically and
it is useful in understanding string theories 
to study noncommutative field theories.

In supersymmetric case,
BPS solitons are important
to investigate non-perturbative properties of 
noncommutative field theories
as in string theories.
Among them,
instantons \cite{NeSc}-\cite{Fu4} and monopoles \cite{HaHa}-\cite{GrNe3}
in noncommutative gauge theories
have been studied intensively.
In string theories,
the instanton and the monopole correspond to
a D(p-4)-brane in a Dp-brane and
a D(p-2)-brane attached to a Dp-brane, respectively.

On the other hand, some exact non-BPS solitons have been found 
in noncommutative field theories.
They play an important role to study the 
condensation of the tachyon in non-BPS branes 
\cite{GoMiSt}-\cite{BaKaMaTa}.
In particular, it was shown that 
a transformation of a solution of the equation of motion
became a new solution of it and
using this solution generating technique, 
the exact solitons were found
in the effective open string field theory \cite{HaKrLa}.
This transformation is almost a gauge transformation 
and is generated by a non-unitary operator $S$,
which satisfies $S^\dagger S=1$ and $ S S^\dagger=1-P$
where $P$ is a projection operator.
This ``gauge'' transformation 
adds localized solitons to the known soliton.
However, we easily see that the BPS equation is not always satisfied 
by the configuration constructed from the BPS soliton by this transformation.
This is because the BPS equation has a constant part.
For example, the transformation of $1+{\cal O}$ 
is $ 1+{\cal O} \rightarrow 1+S {\cal O} S^\dagger 
\neq S(1+{\cal O})S^\dagger$.

In this paper we construct 
new exact BPS solitons in some noncommutative gauge theories
by the ``gauge'' transformation of known BPS solitons.
This can be achieved by the tuning of parameters of theories 
or the modification of the transformation.
Moreover, we add the constant elements to the transformed gauge fields
which show the moduli parameters of the added localized solitons. 
These solitons include the bound state of 
a noncommutative Abelian monopole \cite{GrNe}
and $N$ fluxons \cite{GrNe2} at threshold. 
This corresponds, in superstring theories, to 
a D-string which attaches to a D3-brane 
and $N$ D-strings which pierce the D3-brane, where all D-strings are
parallel to each other.
The moduli parameters correspond to the positions of the $N$ fluxons.

This paper is organized as follows.
In section 2, we briefly review the exact BPS solutions in
noncommutative gauge theories. 
In section 3, we present a BPS-solution generating technique 
and construct new BPS solutions 
in noncommutative gauge theories.
Finally section 4 is devoted to conclusion.

\section{Review of some BPS Solutions}

In this section, 
we shall review some BPS solutions 
in noncommutative gauge theories 
such as vortices \cite{JaMaWa,Ba} and an Abelian monopole \cite{GrNe}.

First we establish notations.
The coordinates obey the following commutation relations
\begin{eqnarray}
\label{noncom}
[x^i,x^j]=i\theta^{ij}.
\end{eqnarray}
Here we consider the case for ${\rm rank} (\theta) =2 $ and 
choose the convention of $\theta^{ij}$ as $\theta^{12}=\theta>0$.
The case for ${\rm rank} (\theta) >2 $ will be discussed later. 

We introduce the complex coordinates as
\begin{eqnarray}
z\equiv x^1+ix^2,~~~\zb\equiv x^1-ix^2.
\end{eqnarray}
Because of the space-space noncommutativity (\ref{noncom}),
we can define annihilation and creation operators in a Fock space $\cH$ as
\begin{eqnarray}
&~&a\equiv\fr{1}{\sqrt{2\theta}}z,~~~a^\dagger\equiv\fr{1}{\sqrt{2\theta}}
\zb,
\end{eqnarray}
so that
\begin{eqnarray}
&~&[a,a^\dagger]=1,~~~[a,a]=[a^\dagger,a^\dagger]=0.
\end{eqnarray}
$\cH$ is spanned by $\vert n\ket, {\rm where} ~a^\dagger a\vert n\ket 
=n\vert n\ket,~n\geq 0$.
Note that the derivative of an arbitrary operator $\phi$
with respect to the noncommutative coordinates $x^i$ 
can be written as 
$\del_i\phi=[\hat{\del}_i,\phi]$
where $ \hat{\del}_i=-i (\theta^{-1})_{ij}x^j$.

Using anti-Hermitian operators 
\begin{eqnarray}
D_i&\equiv& \hat{\del}_i  +A_i,
\end{eqnarray}
in the Fock space,
we define the covariant derivatives as
\begin{eqnarray}
\na_i\phi &\equiv& -\phi \, \hat{\del}_i +D_i\phi,\\
\na_i\Phi &\equiv& [D_i,\Phi],
\end{eqnarray}
where $\phi$ belongs to fundamental representation
and $\Phi$ belongs to adjoint representation
of the noncommutative gauge group.
We can rewrite covariant operators $\na_i$ as
\begin{eqnarray}
\na_i\phi=[\hat{\del}_i,\phi]+A_i\phi.
\end{eqnarray}
If we define
\begin{eqnarray}
D_z= -\frac{1}{\sqrt{2 \theta}} a^\dagger +A_z=\hat{\pa}_z+A_z,
\;\;\,\, D_{\zb}= \frac{1}{\sqrt{2 \theta}} a +A_{\zb}=
\hat{\pa}_{\zb}+A_{\zb}=-D_z^\dagger,
\end{eqnarray}
these are also written by the complex coordinates as
\begin{eqnarray}
\na_z\phi &=& 
-\phi \hat{\pa}_z+D_z \phi,
\hspace{1cm}
\na_{\zb} \phi =
-\phi \hat{\pa}_{\zb}+D_{\zb} \phi, \\
\na_z \Phi &=& [D_z, \Phi],
\hspace{.5cm}
\na_{\zb} \Phi = [D_{\zb}, \Phi],
\end{eqnarray}
where $A_z\equiv \half(A_1-iA_2)$ and 
$A_{\zb}=-A_z^\dagger\equiv \half(A_1+iA_2)$.

The field strength is given by
\begin{eqnarray}
F_{ij} \equiv [D_i,D_j]-i (\theta^{-1})_{ij},
\end{eqnarray}
and we also define the magnetic fields as
$B_i\equiv -\dis\fr{i}{2}\epsilon_{ijk}F^{jk}$
where $i,j,k=1,2,3$.
Using the complex coordinates,
we rewrite these as
\begin{eqnarray}
F_{z\zb}&=&-\left([D_z,D_z^\dagger]+\fr{1}{2\theta}\right)\\
B_z&\equiv&\half(B_1-iB_2),~~~B_{\zb}\equiv\half(B_1+iB_2)\\
B_3&=&2\left([D_z,D_z^\dagger]+\fr{1}{2\theta}\right).
\end{eqnarray}
Note that $\int dx^1 dx^2$ denotes $2\pi\theta{\rm Tr}$, where ${\rm Tr}$
is taken over $\cH$.



Now we consider the BPS vortex solution in $(2+1)$-dimensional 
noncommutative Abelian Higgs model.
The action of this gauge theory is given by 
\begin{eqnarray}
S=-\fr{1}{g^2_{\scr\mbox{YM}}}
\int dt d^2x~\left(-\qua F_{mn}F^{mn}+\vert \na_m \phi\vert^2
+\fr{\beta}{2}(\phi\phi^\dagger -v^2)^2\right),
\label{uzu}
\end{eqnarray}
where $\phi$ is a fundamental scalar field
which is taken so that the coefficient of the kinetic term of 
$\phi$ should be $-1/g^2_{\scr\mbox{YM}}$. 
Here we set the parameter 
$\beta=1$ so as to guarantee BPS condition \cite{JaMaWa}.
The self-dual BPS equations are
\begin{eqnarray}
\label{bogo_B}
&~&B_3=v^2-\phi\phi^\dagger ,\\
\label{bogo_phi}
&~&\na_{\zb} \phi=0,~~~\na_z\phi^\dagger =0.
\end{eqnarray}
The BPS solution for this theory have not been 
found for generic $\theta$.
For the anti-self-dual BPS equations, however, at large $\theta$, 
the solution was derived in \cite{JaMaWa}.
Note that this action and the BPS equations are 
not invariant under the permutation of $\phi$ and $\phi^\dagger$
because of the noncommutativity.
This fact explains why 
the BPS states derived in \cite{JaMaWa}
can not have the negative winding number.
Similar argument holds in $(2+1)$-dimensional
noncommutative pure Yang-Mills model \cite{GrNe3}.


In contrast to the Abelian Higgs model, 
exact BPS solutions in $(3+1)$-dimensional noncommutative Abelian
gauge theory have been obtained in \cite{GrNe} \cite{GrNe2}.
Here we take $x^0,~x^3$ as commutative coordinates and
$x^1,~x^2$ as noncommutative coordinates.
The action is given by
\begin{eqnarray}
S=-\fr{1}{4g^2_{\scr\mbox{YM}}}\int d^4 x~\left(F_{\mn}F^{\mn}
+2\na_\mu\Phi\na_\mu\Phi\right),
\label{abelian4}
\end{eqnarray}
where $\Phi$ is an adjoint Higgs field.
The BPS equations are
\begin{eqnarray}
B_z=\pm\nabla_z\Phi,~~~
B_{\zb}=\pm\nabla_{\zb}\Phi,~~~
B_3=\pm\nabla_3\Phi.
\label{bps1}
\end{eqnarray}
As is found in \cite{GrNe}, the exact one-monopole solution of (\ref{bps1}) 
is 
\begin{eqnarray}
\label{gro-nek-sol}
 \Phi&=&\sum_{n=0}^{\infty}\Phi_n\vert n \ket\bra n \vert
    =\pm\left\{ \sum_{n=1}^{\infty}\left(\xi_n^2-\xi_{n-1}^2\right)
       \vert n \ket\bra n \vert+\left(\xi_0^2+\fr{x_3}{\theta}\right)
       \vert 0 \ket\bra 0 \vert \right\},\no\\
A_z&=&\fr{1}{\sqrt{2\theta}}\sum_{n=0}^{\infty}\left(1-\fr{\xi_n}
       {\xi_{n+1}}\right)a^\dagger\vert n \ket\bra n \vert,~~~
A_3=0,
\end{eqnarray}
where
\begin{eqnarray}
\zeta_n\equiv\int_0^\infty dp~p^ne^{-\theta p^2+2px_3},~~~
\xi_n\equiv\sqrt{\fr{n\zeta_{n-1}}{2\theta\zeta_n}}.
\end{eqnarray}
Here we understand $\xi_0=\sqrt{\fr{1}{2\theta\zeta_0}}$.
The field strength of the solution is 
\begin{eqnarray}
B_z&=&\fr{1}{\sqrt{2\theta}}\sum_{n=0}^{\infty}\fr{\xi_n}{\xi_{n+1}}
       (\Phi_n-\Phi_{n+1})a^\dagger\vert n \ket\bra n \vert,
\no\\
B_3&=&\fr{1}{\theta}\sum_{n=1}^{\infty}\left(1-(n+1)
      \fr{\xi_n^2}{\xi_{n+1}^2}+n
      \fr{\xi_{n-1}^2}{\xi_n^2}\right)\vert n \ket\bra n \vert
      +\fr{1}{\theta}\left(1-\fr{\xi_0^2}{\xi_1^2}\right) 
      \vert 0 \ket\bra 0 \vert.
\end{eqnarray}

We note that the action (\ref{abelian4}) can be regarded as
the effective action on the world volume of a D-brane.
Indeed, it was shown that taking the zero slope limit \cite{SeWi},
the tree-level world volume action of 
Dp-branes in background NS-NS $B$ field becomes 
the $(p+1)$-dimensional noncommutative gauge 
theory with sixteen supersymmetries.
Here the noncommutativity is given by $\theta =1 / B$.
This theory has $9-p$ Higgs fields $\Phi^\mh$
which correspond to the transverse 
coordinates. The action (\ref{abelian4}) is obtained from this 
world volume action by setting $\Phi^\mh=0$ where $\mh=5, \ldots ,9$
and $\Phi^4 \equiv \Phi$.
A monopole solution corresponding to a D(p-2)-brane
is mapped to an anti-monopole solution corresponding to
an anti-D(p-2)-brane by $\Phi \rightarrow -\Phi$.

\section{BPS-Solution Generating Technique and New BPS Solutions}

Now we will construct exact new BPS solutions by transformation of 
BPS solutions. 
The transformation generating operator $S$ is defined as
\begin{eqnarray}
S^\dagger S=1,~~~SS^\dagger=1-P_1,
\end{eqnarray}
where $P_N$ is a projection operator onto $N$-dimensional subspace of
$\cH$ and defined as
\begin{eqnarray}
P_N\equiv \sum_{m=0}^{N-1}\vert m\ket\bra m \vert.
\end{eqnarray}
The almost unitary 
generator $S$ and the projection $P_1$ satisfy the equations
\begin{eqnarray}
P_1 S=S^\dagger P_1=0.
\end{eqnarray}

Up to the noncommutative gauge equivalence,
this operator is represented in the occupation number basis as
\begin{eqnarray}
S&=&\sum_{n=0}^{\infty}\vert n+1\ket\bra n \vert,~~~
S^\dagger = \sum_{n=0}^{\infty}\vert n\ket\bra n+1 \vert.
\end{eqnarray}
We also define the following operator  
\begin{eqnarray}
S_N&\equiv&S^N=\sum_{n=0}^{\infty}\vert n+N\ket\bra n \vert,~~~
S_N^\dagger \equiv (S^\dagger)^N=\sum_{n=0}^{\infty}\vert n\ket\bra n+N \vert,
\end{eqnarray}
which satisfies 
\begin{eqnarray}
S_N^\dagger S_N=1,~~~S_NS_N^\dagger=1-P_N,~~~P_N S_N=S_N^\dagger P_N=0.
\end{eqnarray}

We shall transform the gauge field $A$ (or anti-Hermitian operator $D=d+A$) 
and the Higgs fields $\phi$ and $\Phi$ by the
transformation generating operator $S,S^\dagger$ as
\begin{eqnarray}
\label{gauge_trf}
&~&D_z\rar S_ND_zS_N^\dagger, \,\,\,\,\, \Phi \rar S_N\Phi S_N^\dagger
,\,\,\,\,\,\phi\rar S_N\phi.
\end{eqnarray}
This transformation is similar to the noncommutative 
gauge transformation.
{}From now on, we will call this transformation ``gauge'' transformation.
Note that 
the solution of the equation of motion is
transformed by $S_N$ to another solution of the equation of motion
as was discussed in 
\cite{HaKrLa}. (see also \cite{Ho} \cite{Fu2} \cite{Ne}
\cite{Wi} \cite{AgGoMiSt}.) 

In the following,
we will construct a set of new solutions of the BPS equations, 
instead of the solution of the equations of motion, 
by this ``gauge'' transformation from BPS solutions.
Moreover, we will find that the transformed gauge fields
can have the constant elements which show moduli parameters.


\subsection{New Exact BPS Solution in $(2+1)$-dimensional 
Noncommutative Abelian Higgs Model}
\hspace*{5mm}

Suppose that a set of gauge field $A$ and Higgs field $\phi$ 
is BPS solution in 
$(2+1)$-dimensional noncommutative Abelian Higgs model 
with the action (\ref{uzu}), i.e. it satisfies BPS
equations (\ref{bogo_B}), (\ref{bogo_phi}).
First, let us transform it by the ``gauge'' transformation (\ref{gauge_trf}).
Under this transformation, the left and right hand side of 
the BPS equations (\ref{bogo_B}), (\ref{bogo_phi}) becomes
\begin{eqnarray}
\mbox{l.h.s. of (\ref{bogo_B})}&:&
B_3=2[D_z,D_z^\dagger]+\fr{1}{\theta}\rar 
S_N\left(2[D_z,D_z^\dagger]+\fr{1}{\theta}\right)S_N^\dagger
+\fr{1}{\theta}P_N\\
\mbox{r.h.s. of (\ref{bogo_B})}&:&v^2-\phi\phi^\dagger \rar 
S_N(v^2-\phi\phi^\dagger )S_N^\dagger+v^2 P_N,\\
\mbox{l.h.s. of (\ref{bogo_phi})}&:&\na_{\zb}\phi\rar 
S_N \na_{\zb}\phi,~~~\na_z\phi^\dagger\rar (\na_z\phi^\dagger)S_N^\dagger.
\end{eqnarray}
{}From the above relations, we can find a new BPS solution by the ``gauge'' 
transformation of an original
BPS solution only for $\theta=1/v^2$
, which means that the scale of noncommutativity equals to that of vortices. 
Here we note that BPS equations still remain intact adding the elements
$\sum_{m=0}^{N-1}\lambda_m\vert m\ket\bra m \vert$, where $\lambda_m$
are constants,
to the transformed gauge fields. 
General solution is 
\begin{eqnarray}
D_z^{\scr\mbox{new}}=S_N D_zS_N^\dagger+\sum_{m=0}^{N-1}
\lambda^z_m\vert m\ket\bra m \vert,~~~
\phi^{\scr\mbox{new}}=S_N\phi,
\end{eqnarray}
where $\lambda^z_m$ are arbitrary complex constants and
are interpreted as the positions of the solitons added by the
transformation
\cite{GrNe3}.

The solution constructed from the vacuum state $A_z=0,~\phi=v$ 
is the solution found by Bak \cite{Ba} setting all $\lambda^z_m$ zero:
\begin{eqnarray}
\label{baksol}
\phi=v\sum_{n=0}^{\infty}\vert n+N\ket \bra n\vert,~~~
A_z=-\fr{1}{\sqrt{2\theta}}\sum_{n=0}^{\infty}
\left(1-\sqrt{\fr{n+1-N}{n+1}}\right)a^\dagger\vert n\ket \bra n\vert.
\end{eqnarray}
As long as $\theta=1/v^2$, we can construct various BPS solution
$D_z^{\scr\mbox{new}}$ and $\phi^{\scr\mbox{new}}$ from
arbitrary known BPS solutions besides the vacuum.

\subsection{New Exact BPS Solution in $(3+1)$-dimensional 
Noncommutative Abelian Gauge Theory}

Next, we consider 
$(3+1)$-dimensional noncommutative Abelian gauge theory with 
an adjoint Higgs field.
The action is given by (\ref{abelian4}).
Suppose that a set of gauge fields $A_i$ and adjoint Higgs field $\Phi$ is
BPS solution, i.e. it satisfies BPS equations 
\begin{eqnarray}
\label{bogo_3}
2[D_z,D_z^\dagger]+\fr{1}{\theta}&=&\pm[D_3,\Phi],\\
\label{bogo_a}
[D_3,D_z]&=&\pm[D_z,\Phi].
\end{eqnarray}
Let's transform the gauge fields $A_i$ and adjoint Higgs field $\Phi$ as
\begin{eqnarray}
&~&D_z\rar S_N D_z S_N^\dagger,~~~\Phi\rar S_N\Phi S_N^\dagger,\no\\
&~&D_3=\del_3+A_3 \rar \del_3+S_NA_3S_N^\dagger.
\end{eqnarray}
Note that we do not transform $D_3$ as $D_3 \rar S_N D_3
S_N^\dagger$. $S_N$ does not depend on $x^3$ and satisfies $S_N^\dagger
S_N=1,~S_NS_N^\dagger=1-P_N$.
Under this ``gauge'' transformation, 
the left and right hand side of the BPS equation  becomes
\begin{eqnarray}
\mbox{l.h.s. of (\ref{bogo_3})}&:&
2[D_z,D_z^\dagger]+\fr{1}{\theta}\rar S_N\left(2[D_z,D_z^\dagger]
+\fr{1}{\theta}\right)S_N^\dagger+\fr{1}{\theta}P_N,\\
\mbox{r.h.s. of (\ref{bogo_3})}&:&
[D_3,\Phi]\rar S_N[D_3,\Phi]S_N^\dagger,\\
\mbox{l.h.s. of (\ref{bogo_a})}&:&
[D_3,D_z]\rar S_N[D_3,D_z]S_N^\dagger,\\
\mbox{r.h.s. of (\ref{bogo_a})}&:&
\label{rhs_bogo_a}
[D_z,\Phi]\rar S_N[D_z,\Phi]S_N^\dagger,
\end{eqnarray}
{}From the above relation, we should modify the ``gauge'' transformation for 
$\Phi$,
so that the transformed configuration should be BPS.
The modified ``gauge'' transformation would be 
\begin{eqnarray}
\Phi&\rar&S_N\Phi S_N^\dagger\pm \fr{x^3}{\theta} P_N.
\end{eqnarray}
This transformation leaves the BPS equations (\ref{bogo_3}), 
(\ref{bogo_a}) intact because the transformation of r.h.s. of (\ref{bogo_3})
is modified as $[D_3,\Phi]\rar S_N[D_3,\Phi]S_N^\dagger\pm \fr{1}{\theta}
P_N$ and that of r.h.s. of (\ref{bogo_3}) is the same as (\ref{rhs_bogo_a})
owing to $S_N^\dagger P_N=P_N S_N=0$.
Moreover, as in the case of vortices, 
BPS equations still remain intact adding the constant elements
to the transformed gauge fields. 
General solution is
\begin{eqnarray}
\label{general_Phi}
\Phi^{\scr\mbox{new}}&=& S_N\Phi S_N^\dagger\pm\fr{x^3}{\theta} P_N
+\sum_{m=0}^{N-1}\lambda^4_m
\vert m\ket\bra m \vert ,\\
D_3^{\scr\mbox{new}}&=&\del_3+S_NA_3S_N^\dagger
+\sum_{m=0}^{N-1}\lambda^3_m
\vert m\ket\bra m \vert,\\
\label{general_Dz}
D_z^{\scr\mbox{new}}&=&S_ND_zS_N^\dagger+\sum_{m=0}^{N-1}\lambda^z_m
\vert m\ket\bra m \vert.
\end{eqnarray}
where $\lambda_m^z\equiv \lambda_m^1+i\lambda_m^2$ and 
$\lambda_m^1,~\lambda_m^2,~\lambda_m^3,~\lambda_m^4$ 
are arbitrary real constants.

The solution constructed from the vacuum state $A=0,~\Phi=0$ is 
the $N$-fluxon solution  \cite{GrNe2} 
\begin{eqnarray}
D_z&=&S_N\hat{\del}_z S_N^\dagger,~~~A_3=0,~~~
\Phi= \pm \fr{x^3}{\theta} P_N.
\label{fluxon}
\end{eqnarray}


In \cite{GrNe}, an exact BPS solution 
(\ref{gro-nek-sol}) have been constructed in
noncommutative Abelian gauge theory by Nahm construction.
They also found the solution (\ref{fluxon}) that describes 
infinite D1 strings piercing
a D3 brane, which they call the fluxons \cite{GrNe2}.
Here we construct a new solution by the ``gauge'' transformation from 
the solutions (\ref{gro-nek-sol}).
The solution is following 
\begin{eqnarray}
\Phi^{\scr\mbox{new}}&=&
=\sum_{n=N}^{\infty}\Phi_{n-N}\vert n\ket\bra n \vert+\sum_{m=0}^{N-1}
\left(\pm\fr{x^3}{\theta}+\lambda^4_m\right)
\vert m\ket\bra m \vert, \no\\
D_z^{\scr\mbox{new}}&=&
=\fr{1}{\sqrt{2\theta}}\sum_{n=N}^{\infty}\sqrt{\fr{n+1-N}{n+1}}
\fr{\xi_{n-N}}{\xi_{n+1-N}}a^\dagger\vert n\ket\bra n \vert
+\sum_{m=0}^{N-1}\lambda^z_m
\vert m\ket\bra m \vert,\no\\
A_3^{\scr\mbox{new}}&=&\sum_{m=0}^{N-1}\lambda_m^3
\vert m\ket\bra m \vert.
\end{eqnarray}
This solution can be interpreted as the bound state at
threshold of an
Abelian monopole and $N$ fluxons
(Figure 1).
\begin{figure}
\label{small}
\epsfxsize=100mm
\hspace{3cm} 
\epsffile{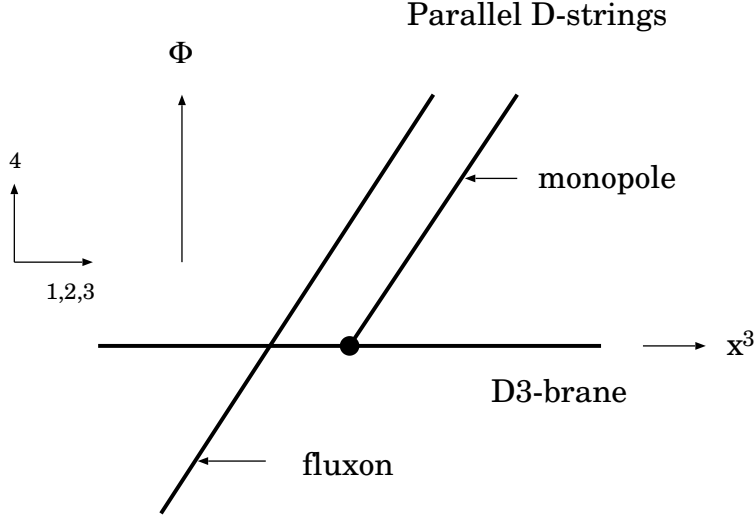}
\caption{Bound state at threshold of an
Abelian monopole and $N$ fluxons ($N$=1)}
\end{figure}  

The new solution can be represented as
\begin{eqnarray}
D^{\scr\mbox{new}}=S_NDS_N^\dagger+\sum_{m=0}^{N-1}\lambda_m \vert m
\ket\bra m\vert=\left(
\begin{array}{ccc|c}
\lambda_0&~&O&~\\
~&\ddots&~&O\\
O&~&\lambda_{N-1}&~\\\hline
~&O&~&D
\end{array}\right),
\end{eqnarray}
where $\lambda_m$ are real constants.
The transformation by $S_N$ corresponds to the shift of the matrix
elements in the lower-right direction by $N$ \cite{AgGoMiSt}. 
The $\lambda_m$ can be interpreted
as the coordinates of localized solitons in matrix theoretical picture
although the action is difficult to 
be realized in the matrix models \cite{BaFi}-\cite{Ao}
because of the commutative coordinates $x^0$ and $x^3$.
In this monopole case, the localized solitons are
fluxons. This picture is also applicable to vortices and instantons.

We have set the transverse coordinates $\Phi^\mh=0$ 
in the last paragraph in section 2.
After the transformation, however, we can take $\Phi^\mh\neq 0$ 
keeping the BPS condition.
For example, to the general solutions
(\ref{general_Phi})-(\ref{general_Dz}), we can set
\begin{eqnarray}
\label{transverse}
\Phi^\mh=\sum_{m=0}^{N-1}\lambda_m^\mh \vert m\ket\bra m\vert=\left(
\begin{array}{ccc|c}
\lambda_0^\mh&~&O&~\\
~&\ddots&~&O\\
O&~&\lambda_{N-1}^\mh&~\\\hline
~&O&~&O
\end{array}\right),
\end{eqnarray}
where $\lambda_m^\mh,~\mh=5,\ldots,9$ are real constants and 
denote the $\mh$-th transverse
coordinates of the $m$-th fluxon. This shows that the $N$ fluxons can 
escape from the D3-brane
(Figure 2).
\begin{figure}
\epsfxsize=120mm
\label{small2}
\hspace{2cm}
\epsffile{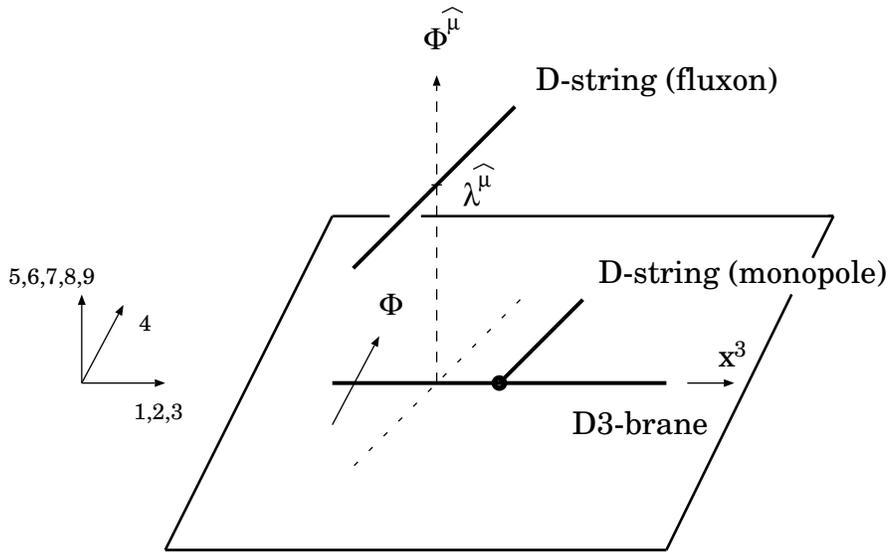}
\caption{Fluxons can escape from the D3-brane.}
\end{figure}  

\subsection{Exact BPS Instanton Solution in $4$-dimensional Noncommutative
Gauge Theory}

As in the previous case, 
we can also obtain the 
exact BPS solutions in $4$-dimensional Euclidean
noncommutative gauge theory
with the action 
\begin{eqnarray}
S=-\fr{1}{4g^2_{\scr\mbox{YM}}}\int d^4x~\Tr F_{\mn}F^{\mn}.
\end{eqnarray}
The BPS equations are
\begin{eqnarray}
F_{z_1\zb_1} \pm F_{z_2\zb_2}=0,~~~F_{z_1z_2}=0,
\end{eqnarray}
where $z_1\equiv x^1+ix^2,~z_2\equiv x^3+ix^4$. 

Here we restrict ourselves to 
consider the gauge theory with 
the anti-self-dual noncommutative parameter $\theta^{\mn}$.
In this case, we can take $\theta^{12}=\theta>0, \theta^{34}=-\theta$
and (other components) $=0$ without loss of generality.
We define anti-Hermitian operators as 
\begin{eqnarray} 
D_{z_i} &\equiv& -\fr{1}{2\theta} \zb_i +A_{z_i}
,
\end{eqnarray}
where $i=1,2$ and 
$A_{z_1}\equiv \half(A_1-iA_2),~A_{z_2}\equiv \half(A_3-iA_4)$.

For the anti-self-dual $\theta$,
the self-dual noncommutative instantons were investigated intensively.
In particular, their moduli space is resolved by noncommutativity.
Thus the instanton can not escape from the brane and
we can construct the instanton even in the commutative $U(1)$ gauge
theory with DBI action
\cite{SeWi} \cite{Te}.
However, the anti-self-dual noncommutative instanton has
the same moduli space as commutative one which has small instanton
singularities.
The solution which can be obtained by the ``gauge'' transformation
is an anti-self-dual BPS solution \cite{AgGoMiSt}
since the added solitons can escape from the brane.
Thus we concentrate on the anti-self-dual 
BPS solution.

Suppose that $A_\mu$ is an anti-self-dual 
BPS solution in 4-dimensional noncommutative  
gauge theory, i.e. it satisfies the anti-self-dual BPS equation 
\begin{eqnarray}
\label{asd}
[D_{z_1},D^\dagger_{z_1}]+[D_{z_2},D^\dagger_{z_2}]=0,~~~
[D_{z_1},D_{z_2}]=0.
\label{asbps}
\end{eqnarray}
We note that the constants terms in 
\begin{eqnarray} 
F_{z_i\zb_i}=-[D_{z_i},D_{z_i}^\dagger]+\frac{1}{2 \theta} (-1)^{i},
\end{eqnarray}
are canceled in the anti-self-dual BPS equation (\ref{asbps}).

In this instanton case, 
the state of the Fock space $\cH_1\otimes\cH_2$ is labeled 
by two non-negative integer numbers e.g. $\vert n_1,n_2\ket$. 
Hence we shall introduce 
a transformation generating operator $T$ \cite{AgGoMiSt} as
\begin{eqnarray}
T^\dagger T=1,~~~TT^\dagger=1-P,
\end{eqnarray}
where $P$ is a projection operator in the Fock space $\cH_1\otimes\cH_2$
which project onto the finite dimensional subspace of
$\cH_1\otimes\cH_2$.
We can show $PT=(1-T T^\dagger)T=0$.
Using this almost unitary operator $T$, we shall transform 
$D$ as before
\begin{eqnarray}
\label{gauge_trf2}
D_{z_i}\rar TD_{z_i}T^\dagger
.
\end{eqnarray}
Since 
\begin{eqnarray}
[D_{z_i},D_{z_j}]&\rar& T[D_{z_i},D_{z_j}]T^\dagger, \\
{[D_{z_i},D^\dagger_{z_j}]}&\rar& T[D_{z_i},D^\dagger_{z_j}]T^\dagger,
\end{eqnarray}
we can construct a new BPS solution from the original BPS solution
by the transformation (\ref{gauge_trf2}).
As in the previous case, 
we introduce 
a projection $P_m$ onto the single state 
which satisfies ${\rm Tr} P_m=1$ and $P=\sum_{m=0}^{N-1}P_m$.
Then general solution constructed by this way can be written as 
\begin{eqnarray}
D_{z_i}^{\scr\mbox{new}}=TD_{z_i}T^\dagger
+\sum_{m=0}^{N-1}\lambda^i_m P_m,
\label{nass}
\end{eqnarray}
where $i=1,2$ and $\lambda^i_m$ are arbitrary complex constants.
As in the monopole case,
we can also set $\Phi^\mh$ as (\ref{transverse}). The localized solitons
are identified as small instantons. This shows that
the $N$ small instantons which correspond to $N$ D(p-4)-branes can escape
from the Dp-brane. 

The solution (\ref{nass}) constructed from the vacuum state $A=0$ 
was derived in \cite{AgGoMiSt}.
As long as the self-duality of the gauge fields is the same as
that of noncommutative parameters $\theta^{\mn}$,
we can construct various BPS solution
from arbitrary known BPS solutions besides the vacuum,
e.g. from $U(2)$ 1-instanton solution in \cite{Fu4}.
We can see that the BPS solution 
constructed from the BPS solution
with the instanton number $M$ 
has the instanton number $M+N$ where $N={\Tr}P$
because of the properties of the projection.

\section{Conclusion}

In this paper, we have constructed 
exact BPS solitons in various noncommutative gauge theories
by the ``gauge'' transformation of known BPS solitons.
These solutions have appropriate physical interpretations,
for example, we have found the bound state of 
a noncommutative Abelian monopole and $N$ fluxons at threshold. 
This corresponds, in superstring theories, to 
a D-string which attaches to a D3-brane 
and $N$ D-strings which pierce the D3-brane, where all D-strings are
parallel to each other.

If we treat non-Abelian gauge theories, we should add Chan-Paton index
$i$ to the state, i.e. the state is written as $\vert n,i\ket$
. Then it is trivial to extend the solution generating technique
to the non-Abelian gauge theory and
we can use the $U(2)$ monopole solution \cite{GrNe3} and
$U(2)$ anti-self-dual instanton solution \cite{Fu4}.
The BPS solution for a non-Abelian Higgs model can be also obtained
from the vacuum.

The application for another noncommutative BPS solitons 
remains to be investigated.
In $\C P_n$ model on noncommutative plane \cite{LeLeYa}, 
it seems to be difficult to generate another 
BPS solution from a BPS solution except for the solution from the vacuum.
It is interesting to study which noncommutative theories 
our technique would be to apply to.

\vskip7mm\noindent
{\bf Acknowledgements}

We would like to thank Y. Matsuo for useful comments and encouragement. 
We are also grateful to
K. Furuuchi, K. Hosomichi, T. Kawano, T. Takayanagi and T. Uesugi
for discussion.
The work of S.T. was supported in part by JSPS Research Fellowships 
for Young Scientists.
The work of M.H. was supported in part by the Japan Securities 
Scholarship Foundation (\#12-3-0403).

\vskip7mm\noindent
{\bf Note Added}

After submitting this paper, we received the paper \cite{Ha}
which partially overlaps our results.

\end{document}